# The Ground state of $BiFeO_3$: Low temperature magnetic phase transitions revisited

Arun Kumar and Dhananjai Pandey[a)]
School of Materials Science and Technology, Indian Institute of Technology (Banaras Hindu University), Varanasi-221005, India.

## ABSTRACT

Recent neutron diffraction and NMR studies suggest that the incommensurately modulated spin cycloid structure of $BiFeO_3$ is stable down to 4.2 K, whereas DC [M(T)] and AC [$\chi$ ($\omega$, T)] magnetization, and caloric studies have revealed several magnetic transitions including a spin glass transition around 25 K. The two sets of observations are irreconcilable and to settle this, it is important to first verify if the low temperature magnetic transitions are intrinsic to $BiFeO_3$ or some of them are offshoots of oxygen vacancies and the associated redox reaction involving conversion of $Fe^{3+}$ to $Fe^{2+}$. We present here the results of M (T) and $\chi$ ($\omega$, T) measurements on pure and 0.3 wt% $MnO_2$ doped $BiFeO_3$ samples in the 2 to 300 K temperature range. It is shown that $MnO_2$ doping increases the resistivity of the samples by three orders of magnitude as a result of reduced oxygen vacancy concentration. A comparative study of the M (T) and AC $\chi$ ($\omega$, T) results on two types of samples reveals that the transitions around 25 K, 110 K and 250 K may be intrinsic to $BiFeO_3$. The widely reported transition at 50 K is argued to be defect induced, as it is absent in the doped samples. We also show that the spin glass transition temperature $T_{SG}$ is less than the spin glass freezing temperature ($T_f$), as expected for both canonical and cluster glasses, in marked contrast to an earlier report of $T_{SG}$ > $T_f$ which is unphysical. We have also observed a cusp corresponding to the spin glass freezing at $T_f$ in ZFC M (T) data not observed so far by previous workers. We argue that the ground state of $BiFeO_3$ consists of the coexistence of the spin glass phase with the long range ordered AFM phase with a cycloidal spin structure.

## I. INTRODUCTION

Bismuth ferrite ($BiFeO_3$) is one of the most investigated magneto-electric multiferroics because of its room temperature multiferroicity with potential for applications in multifunctional devices of technological importance.[1-6] The room temperature ferroelectric phase of bulk $BiFeO_3$ corresponds to a rhombohedrally distorted perovskite structure in the *R*3c space group[7] in which the cations are displaced with respect to the anions along the [111] direction while the neighbouring oxygen octahedra are rotated in opposite directions (antiphase tilted structure in the $a^-a^-a^-$ tilt system[8]) about the same direction. The ferroelectric phase transition temperature is reported to be $T_C$ ~ 1103K.[9] The magnetic structure of $BiFeO_3$ corresponds to a non-collinear *G*-type antiferromagnetic ordering with a superimposed incommensurate magnetic modulation below the Neel temperature $T_N$ ~ 643K.[10] While the nuclear and magnetic structures as well as the multiferroic properties of $BiFeO_3$ at and above the room temperature are well settled[4], there exists considerable controversy about its true ground state. Recent NMR studies suggest that the cycloidal modulation function for the magnetic phase changes from harmonic (sinusoidal) to anhormonic (sn(x, m), elliptic Jacobi function) with m, which is a measure of anhormonicity, increasing from 0.48 at room temperature to 0.95 at 4.2 K[11-14]. The neutron diffraction studies, which probe the space and time averaged magnetic structure at the bulk level, have also confirmed anhormonic nature of modulation of the cycloid at low temperatures[15,16] but the anhormonicity is found to be much less, 0.50[15] and <0.25[16] in two independent studies using single crystals and polycrystalline samples, respectively, than that reported using a local probe like NMR. However, it is hard to imagine that the spin cycloid involving all the spins in the magnetic structure of $BiFeO_3$ will remain unaffected despite the several low temperature phase transitions that have been reported using a variety of measurement probes like DC magnetic susceptibility ($\chi_{DC}$),[17-22] AC magnetic susceptibility [ $\chi'_{AC}$ $(\omega, T)$ and $\chi''_{AC}(\omega, T)$ ],[17,23] differential scanning calorimetry (DSC)[18]/

differential thermal analysis (DTA),[24] Raman scattering[25-31] and elastic modulus spectroscopy.[24,31] Towards understanding the true ground state of $BiFeO_3$, it is therefore important to settle if the low temperature phase transitions reported below room temperature are intrinsic to $BiFeO_3$ or some of them could be induced due to the presence of ionic vacancies created during high temperature processing. The present work was undertaken to address this issue.

Before presenting our results, we first briefly review the present status of the understanding of the low temperature phase transitions in $BiFeO_3$ and associated controversies. Historically, a frequency dependent anomaly in the temperature dependence of AC magnetic susceptibility well below the room temperature was first reported by Nakamura et al.[23] using splat quenched amorphous $BiFeO_3$ samples. Their results showed a cusp in $\chi'_{AC}\ (\omega, T)$ around $T_f$ ~21 K with a spin glass transition temperature $T_{VF}$ ~14 K. Splat quenching was used to suppress the $Fe^{3+}$ to $Fe^{2+}$ redox reaction, supposedly without affecting the magnetic ordering behaviour of $BiFeO_3$, which occurs due to the electrons released by the creation of oxygen vacancies in polycrystalline $BiFeO_3$ samples synthesized and sintered at high temperatures. Recent $\chi_{AC}\ (\omega, T)$ measurements on single crystals of $BiFeO_3$ have also revealed frequency dependent cusps around $T_f$ ~25 K[17] similar to that reported by Nakamura et al. though with a slightly higher cusp temperature that may possibly be due to crystalline nature of these samples. However, the spin glass transition temperature $T_{SG}$, which represents the critical slowing down of the spin dynamics, is reported to be ~29.4 K.[17] The $T_{SG}$ in the canonical and cluster spin glasses is known to be lower than the cusp temperature ($T_f$) corresponding to the lowest frequency of measurements and represents the divergence of the time scale associated with the slowest spin dynamics.[32] Since the $T_{SG}$ ~29.4 K reported by Singh et al. is greater than $T_f$ ~25 K corresponding to the cusp temperature for their lowest frequency data (10 Hz), it calls for further investigation.

Besides the $\chi_{AC}$ (ω, T) studies, several new low temperature magnetic transitions have been reported using zero field cooled (ZFC) and field cooled (FC) magnetization [M (T)] measurements in recent years.[17-22] None of these measurements, however, reveal any anomaly around 21-25 K at which AC susceptibility studies reveal spin glass freezing. ZFC M (T) measurements on polycrystalline powders of $BiFeO_3$ have revealed anomalies around 50 K,[19] 150 K[18] and 250 K.[18] Further, the ZFC and FC M (T) curves have been reported to bifurcate below 300 K but the anomaly around 50 K does not appear under the FC condition. This was taken as evidence for spin glass freezing (SG) or superparamagnetic (SPM) blocking. However, no corresponding anomaly is observed in $\chi'_{AC}(\omega, T)$ or $\chi''_{AC}(\omega, T)$ and therefore the possibility of this anomaly being linked with SG freezing or SPM blocking becomes remote. The anomaly around 50 K has also been reported in the ZFC M (T) measurements on single crystals[17] and nanocrystalline powders[19,20] with a bifurcation temperature of ~ 250 K[17] and just below 300 K,[19,20] respectively. There are reports of anomalies around 65 K and 70 K as well but these have been subsequently found to be due to the presence of $Bi_2Fe_4O_9$ and $\gamma$-$Fe_2O_3$ impurity phases, respectively.[21,22]

DSC measurements on polycrystalline $BiFeO_3$ have also revealed a strong specific heat anomaly at ~250 K similar to that observed in ZFC M (T) data[18] whereas the DTA analysis on $BiFeO_3$ single crystals suggests the possibility of a first order phase transition with a latent heat close to 150 K (~140 K).[24] Temperature dependence of the magnetic entropy, on the other hand, indicates five phase transitions occurring around 38 K, 150 K, 178 K, 223 K and 250K.[18] The dielectric measurements reveal weak anomalies around 55 K,[24] 140 K[24] and 215 K[24] for single crystals and around 25 K,[25] 50 K,[24] 200 K[24] and 220-260 K[24,33] in ceramics. Elastic Modulus measurements reveal anomalies around 140 K[24,31] and 200 K,[24] respectively. Further, Raman spectroscopic studies on polycrystalline and single crystals of $BiFeO_3$ have revealed transitions at 25 K,[25] 90 K,[27] 140 K,[26-31] 200 K[26-28] and 250 K.[27] Taking into account the observations made by different experimental probes discussed above, the transition occurring around 38-50 K, 140–150 K and 220–250 K have been attributed to magnetic but glassy with magnetoelectric coupling, predominantly magnetic transition involving spin reorientation, and antiferromagnetic (AFM) to spin glass (SG) transition with weak coupling with polarization, respectively.[18, 24] The transition reported around 178 – 200 K has been proposed to be magnetoelastic in nature with a small coupling to polarization.[24]

It is, however, quite intriguing to note that no single research group has reported all the low temperature transitions in the same sample. The fact that Nakamura et al. did not find any cusp in $\chi_{AC}$ around $T_f$ ~ 50 K or signatures of other transitions above 50 K in splat quenched $BiFeO_3$ samples, supposed to be free from oxygen vacancies, also raises doubts about the intrinsic nature of various low temperature transitions other than the transition occurring around 25 K. Surprisingly, the low temperature M (T) measurements by various workers have not revealed the occurrence of the 25 K transition and its existence has been identified by $\chi_{AC}$ measurements,[17] dielectric measurements[25] and Raman studies[25] only. To settle whether the various low temperature transitions reported by different workers are intrinsic to $BiFeO_3$ or some of them could be linked with cation/anion vacancies, we have carried out a comparative study of the low temperature phase transition behavior of polycrystalline $BiFeO_3$ samples with and without 0.3 wt% $MnO_2$ doping prepared under identical heat treatment conditions using DC magnetization [M (T)], and AC susceptibility [$\chi_{AC}$ (ω, T)] measurements. 0.3 wt% $MnO_2$ doped $BiFeO_3$ samples have been used to understand the role of oxygen vacancies since such a doping is known to decrease the oxygen vacancy concentration, enhance DC resistance of $BiFeO_3$ based ceramics and gives better quality P-E hysteresis loops.[34] Our results provide unambiguous evidence for spin glass freezing around $T_f$ ~25 K in M (T) as well as $\chi_{AC}$ (ω, T) in both the doped and undoped samples with a spin glass transition temperature $T_{SG}$ ~20 K which, unlike the previous report,[23] is less than $T_f$, as expected for such glassy transitions. Our results also suggest that the anomaly around 50 K could be due to extrinsic factors like oxygen vacancies as it is not present in the $MnO_2$ doped samples. The two other anomalies seen in the M (T) of the doped sample seem to suggest that 250 K and 100-150 K transitions reported in the literature could also be intrinsic though it requires more investigation. Our results, in conjunction with the recent observations based on neutron and NMR studies, suggest that the ground state of $BiFeO_3$ corresponds to a coexistence of spin glass phase with the long range spin cycloid. We also discuss the implications of such a phase coexistence in terms of the existing theoretical models.

## II. EXPERIMENTAL DETAILS

**A. Sample Preparation**: Both undoped and 0.3 wt% $MnO_2$-doped $BiFeO_3$ powders were prepared by conventional solid state thermochemical reaction method using high purity oxides as starting materials: $Bi_2O_3$ (Aldrich, 99.9 %), $Fe_2O_3$ (Aldrich, 99 %) and $MnO_2$ (Alfa Aesar, 99.9%). We have added 0.3 wt% $MnO_2$ during calcination for the preparation of doped samples. The ingredient powders were taken in stoichiometric proportions and mixed in an agate mortar pestle for 3 hours and subsequently in a planetary ball mill using a zirconia zar and zirconia balls for 6 hours with acetone as the mixing media. After drying, the mixed powders were calcined at an optimized temperature of 1063 K for 8 hours in an open alumina crucible. The calcined powder was crushed and again ball milled for 4 hours. The powder was then mixed with 2 % poly vinyl alcohol (PVA) as a binder and pressed at a load of 70 KN into disks of 13mm diameter and about 1mm thickness in a hydraulic press using steel die. After binder burn-off at 773 K for 10 hours, sintering were carried out at 1073 K for 1 hour in closed alumina crucibles with calcined powder of the same composition as spacer powder for preventing the loss of $Bi_2O_3$ during sintering. The weight loss during sintering was less than 1%.

**B. Characterizations:** X-ray diffraction (XRD) measurements were carried out using an 18 kW copper rotating anode based (Rigaku, Japan) powder diffractometer operating in the Bragg-Brentano geometry fitted with a curved crystal monochromator in the diffraction beam. The data were collected in the 2θ range of 20 to 120 degrees at a step of 0.02 degrees. The XRD patterns were recorded from powders obtained after crushing the ceramic pellets and annealing the crushed powder at 773 K for 10 hrs for removal of stresses introduced during crushing. The DC resistivity of samples was measured using an electrometer (Keithley model no. 6517A). The dc magnetization measurements were carried out using a VSM (Quantum Design, PPMS) and MPMS SQUID (Quantum Design, MPMS-3) magnetometer in zero-field-cooled (ZFC) and field-cooled (FC) conditions from 2 K to 300 K using dc field of 500 Oe, 20000 Oe and 50000 Oe. For the ZFC measurements, the sample was first cooled from room temperature down to 2 K in the absence of a magnetic field. After applying the magnetic field at 2 K, the magnetization was measured in warming cycle. For the FC measurements, the sample was cooled from room temperature down to 2 K in the presence of magnetic field and magnetization was measured in warming cycle under the same field. The temperature dependent ac magnetic susceptibility measurements were carried out in an MPMS SQUID (Quantum Design MPMS-3) system using AC drive field of 5 Oe at various frequencies ranging from 97.3 Hz to 547.3 Hz. The chemical composition of the pure and 0.3wt% $MnO_2$ doped $BiFeO_3$ samples was checked by Energy Dispersive X-ray spectroscopic (EDX) technique using Carl-Zeiss Scanning Electron Microscope (SEM) model no. EVO 18. The composition was checked over individual grains and around grain boundaries separately. Figs S1 and S2 of the supplementary file give the microstructure and the EDX spectra for the pure and 0.3wt% $MnO_2$ doped samples. The representative regions in the grain and at the grain boundaries are marked in the microstructures. The results of EDX analysis for the two types of samples are given in Table S1. Similar analyses were carried out at five randomly selected regions and the average composition is given in Tables S2 for the pure and 0.3wt% $MnO_2$ doped samples. It is evident from these Tables that the average composition obtained by EDX analysis is close to the nominal (expected) composition within the standard deviation for Bi, Fe and Mn. The oxygen content for the pure sample is found to be a little less than that for the 0.3wt% $MnO_2$ doped sample. However, it is worth mentioning that EDX is not the ideal tool for the determination of oxygen content.

## III. RESULTS AND ANALYSIS

### A. Room Temperature Crystal Structure:

Single-phase powders and sintered ceramic samples of a $BiFeO_3$ are rather difficult to prepare because of the narrow temperature range of the stability of the perovskite phase[36-38] and the volatile nature of $Bi^{3+}$ that promotes the formation of impurity phases like $Bi_2Fe_4O_9$ and $Bi_{25}FeO_{39}$.[37,38] Fig.1. shows the room-temperature x-ray powder diffraction (XRD) patterns of the undoped and 0.3 wt% $MnO_2$ doped $BiFeO_3$ powders. It is evident from the figure that both the samples are almost single phase as all the peaks correspond to the main perovskite phase of $BiFeO_3$ with only a trace amount of an impurity phase $Bi_2Fe_4O_9$ present with a peak intensity that is 1.4 % of the strongest 110 peak of $BiFeO_3$. It is evident from this figure that the $400_{pc}$ is a singlet while $222_{pc}$ and $440_{pc}$ are doublets with the weaker reflection occurring on the lower 2θ side, as expected for the stable rhombohedral phase in the R3c space group of $BiFeO_3$. This was further confirmed through Le-Bail profile fitting for the two samples (see supplementary file for more details). The refined lattice parameters and volume, as given in the Table S3 of the supplementary file, are in good agreement with those reported in the literature. The important inference is that the lattice parameters and volume are not affected by 0.3wt % $MnO_2$ doping.

### B. DC Magnetization Studies:

Figs. 2. [(a), (b), and (c)], and Figs. 3. [(a), (b), and (c)] show the temperature dependent dc magnetization M (T) plots under ZFC and FC conditions for the undoped and doped samples for applied fields of 500, 20000 and 50000 Oe. The ZFC and FC magnetizations bifurcate well above 300K for the undoped samples whereas for the doped samples it begins around 270K at 500 Oe field. The irreversibility of the M (T) plots under ZFC and FC conditions increases with decreasing temperature suggesting spin-glass or superparamagnetic behavior in $BiFeO_3$ at low temperatures as proposed by previous workers also.[17,23] The ZFC and FC magnetization curves of both the samples in the 300 to ~150 K range are consistent with conventional antiferromagnetic (AFM) behaviour in which magnetization is expected to decrease with decreasing temperature below $T_N$. However, both the samples show an upturn in magnetization below ~ 150K. This can happen as a precursor effect to impending phase transition(s) and accordingly a spin reorientation transition[26,27,29] around 150-200K similar to that in orthoferrites,[39,40] with the involvement of electromagnons,[26,29] has been proposed.[17] Such a spin reorientation transition has recently been reported in monoclinic compositions of solid solutions of $BiFeO_3$ with $PbTiO_3$ but the transition temperature is well above the room temperature and just below the $T_N$.[41] In pure $BiFeO_3$, the situation is quite different as spin reorientation has been proposed well below room temperature. However, the ZFC M (T) plot of undoped sample (Fig. 2) does not show any anomaly in the 150 to 200K range, other than the upturn in magnetization around 150K. The corresponding ZFC plot for the doped sample (Fig. 3), however, shows one weak cusp around 110 K. We believe that the presence of vacancy defects in the undoped samples masks this weak anomaly, even though there are indirect indications of such a transition in the undoped sample as well through the upturn in ZFC M (T) and observation of electromagnons.[26,29] What is, however, more significant is the presence of two additional cusps occurring below 110 K, one around 50 K and another around 25 K, in the ZFC M[T] of the undoped sample, whereas the corresponding ZFC plot for the doped sample

shows only one prominent cusp around 25K. The cusp around 25K in ZFC M (T) has not been observed in previous studies on single crystal[17] and polycrystalline[18,19] samples including the early work of Nakamura et al.[23] This clearly underlines the significance of quality and stoichiometry of the sample used in the present investigation in revealing such weak transitions below room temperature and a lack of this is probably the reason why no single worker has so far observed all the low temperature phase transitions in $BiFeO_3$. The fact that the FC M (T) does not show the 25K cusp in both the samples suggests that it could be due to spin glass freezing or superparamagnetic blocking or pinning of a magnetic impurity phase. Further, the absence of the cusp around 50 K in the doped samples suggests that it is a defect (vacancies) induced transition[42-44] that gets suppressed due to $MnO_2$ doping as explained later in this paper. It is interesting to note that the 50 K cusp is observed for the undoped samples under both the ZFC and FC conditions for 500 Oe field suggesting that the spin dynamics associated with this defect-induced cusp is too fast to be frozen or blocked at 500 Oe. On increasing the field strength to 20000 Oe, the 50 K cusp becomes less prominent and it completely disappears in the 50000 Oe field measurement. This suggests that this transition could be associated with faster spin dynamics as compared to the 25K transition and requires higher field to freeze or block the spins. Below 10 K, both ZFC and FC curves show increase in the magnetization indicating a weak ferromagnetic behavior of $BiFeO_3$ at such low temperatures, as noted by previous workers also.[17,18] Since the magnetization values (0.001emu/g) associated with the ferromagnetic rise of M (T) are very small, it is most likely associated with some trace amount of impurities.[4] The M (T) plots of doped $BiFeO_3$ show an additional cusp, though very much less prominent as compared to that at $T_f$, around 260 K in the ZFC and FC conditions. This cusp is not seen in the corresponding plots for the undoped samples. Since doping is expected to suppress the oxygen vacancy concentration, as explained in more detail later on, we believe that the cusps around 25, 110 and 260 K observed in the ZFC M (T) plot of the doped sample are intrinsic to $BiFeO_3$. The previous workers, except Ref,[18] did not observe the anomaly 260 K in the ZFC M (T) plots.[17,19-22]

### C. AC Susceptibility Studies:

In order to determine whether the cusps observed in the M (T) plots below 50 K are due to SG freezing or SPM blocking, we investigated the temperature dependence of the spin relaxation time $\tau$ from AC susceptibility ($\chi\ (\omega, T)$) measurements. The spin relaxation time $\tau$ is the dynamical fluctuation time scale corresponding to the measurement frequency at the peak temperature of the real part of AC magnetic susceptibility. The spin relaxation time (τ) for an assembly of non-interacting superparamagnetic particles slows down gradually as per the Arrhenius law:[32,45]

$$\tau = \tau_0 \exp\left(\frac{E_a}{k_{\mathbf{B}} T}\right) \qquad (1)$$

where $E_a$ is the anisotropy energy barrier equal to KV, where K is the anisotropy energy constant and V is the volume of the particles. In the spin glass systems, on the other hand, one observes critical slowing down of the spin dynamics at a characteristic spin glass transition temperature $T_{SG}$ at which there is ergodicity breaking. In the canonical spin glass systems, an empirical Vogel-Fulcher (VF) law is used to model the critical slowing down of the spins:[32,45]

$$\tau = \tau_0 \exp\left(\frac{E_a}{k_{\mathbf{B}} (T - T_{VF})}\right) \qquad (2)$$

where $\tau_0$ is the time constant corresponding to the attempt frequency, $k_B$ is the Boltzmann constant, $E_a$ is the thermal activation energy and $T_{VF}$ is the Vogel-Fulcher freezing temperature ($0 < T_{VF} < T_f$) like $T_{SG}$. In addition to VF law, a scaling law has also been proposed to characterize the SG transition. The scaling hypothesis assumes that the relaxation time (τ) is related to the correlation length (ξ) near the SG transition temperature ($T_{SG}$). As ξ diverges at $T_{SG}$, relaxation time also diverges[32] as:

$$\tau = \tau_0 \exp\left( \frac{T_f}{T_{SG}} - 1 \right)^{-z\nu} \tag{3}$$

where $\tau_0$ is the characteristic time scale for the spin dynamics, $T_f$ is the maximum temperature corresponding to peak in the real part of $\chi'_{AC}\ (\omega, T)$, $T_{SG}$ is the spin glass transition temperature in the limit of zero frequency, z is the dynamic scaling exponent, and $\nu$ is the critical exponent.

Fig. 4. and Fig. 5. show the real $[\chi'_{AC}\ (\omega, T)]$ and imaginary $[\chi''_{AC}\ (\omega, T)]$ parts of AC magnetic susceptibility measured with an applied ac drive field of 5 Oe in the frequency range 97.3 Hz to 547.3 Hz for undoped and doped $BiFeO_3$ respectively. It is evident from the figures that the peak temperatures corresponding to the real and imaginary parts of AC magnetic susceptibilities shift towards higher temperatures with increasing frequency which can be due to spin glass freezing or SPM blocking. For SPM blocking, ln $\tau$ versus 1/T plot should be linear as per Eq. 1. The non-linear nature of the plots shown in Fig. 6. [(a), (b)] thus rules out SPM blocking in $BiFeO_3$ at low temperatures. Vogel-Fulcher law,[32] on the other hand, provides excellent fit for both the undoped and doped samples, as can be seen from Fig. 6. [(a), (b)], confirming the spin glass freezing with a spin glass transition temperature $T_{SG}$ ~ $T_{VF}$ ~20 K. Attempts to fit the power law type behavior given by Eq. 3 gave inferior fits as shown in Fig. S4 [(a), (b)] of the supplementary file. In fact, the power–law type critical dynamics misses the low frequency $T_f$ data point and the best fit shown in the figures given in the supplementary file after excluding the low frequency data point is still inferior to that obtained by V-F law. The results of the best fit for V-F and power law behaviours are given in Table I. Interestingly, the value of the $T_{SG}$ obtained by power law fit is very close to $T_{VF}$. Also, the exponent ($z\nu$~ $1.5 - 1.8$) is somewhat higher than that reported by Singh et al.[17] for single crystal

$BiFeO_3$ and closer to the hydrodynamic model or mean field model for which $z\nu \sim 2$.[46] The activation energy $E_{act}$ for the V-F law (0.8 meV or about 9.5 K) is comparable to the activation energies reported for the $Eu_{1-x}Sr_xS$ system but lower than those for the canonical RKKY type spin glass systems. Our results thus confirm unambiguously that the anomaly observed around $T_f \sim 25$ K in $BiFeO_3$ is due to spin glass freezing with a spin glass transition temperature $T_{SG}$ ~20 K at which the ergodicity is broken. Our results also show that the earlier report[17] of $T_{SG} > T_f$ in $BiFeO_3$ is not correct and may be an artifact of numerical fit. By definition also, the $T_{SG}$ cannot be higher than $T_f(\omega)$ as it corresponds to the value $T_f(\omega)$ in the limit of ω tending towards zero at which the slowest spin dynamics diverges.

## IV. DISCUSSION OF RESULTS

### A. Role of $MnO_2$ doping:

It is well known that the electrical and magnetic properties of pure $BiFeO_3$ are strongly influenced by oxygen vacancies created during sintering of $BiFeO_3$. Each oxygen vacancy leaves behind two electrons as per the following reaction written in the Kröger-Vink notation:[47]

$$O_0 \Leftrightarrow 1/2\, O_2 + V_0^{\cdot\cdot} + 2e$$

Electrons released due to oxygen vacancy may be captured by $Fe^{3+}$ of $BiFeO_3$ leading to its reduction to $Fe^{2+}$:

$$Fe^{3+} + 1e \Leftrightarrow Fe^{2+}$$

The presence of $Fe^{2+}$ and $Fe^{3+}$ ions leads to hopping of electrons due to which conductivity increases. Poor insulating resistance masks the observation of intrinsic ferroelectric polarization through the P-E hysteresis loop measurements and therefore such samples are not desirable for multiferroic applications. The $Fe^{3+}$ to $Fe^{2+}$ redox reaction also raises the possibility of creating local ferromagnetic clusters[48] via double exchange mechanism and may thus enhance the magnetization.[49-53] Theoretically, the role of intrinsic point defects, especially oxygen vacancies, as a possible source of magnetization in $BiFeO_3$[54] has been quite controversial. Ederer and Spaldin investigated the effect of oxygen vacancies on the weak ferromagnetism of $BiFeO_3$ using first principle calculations.[55] They found that oxygen vacancies lead to the formation of $Fe^{2+}$ which can be identified by the clear qualitative differences in the local density of states, but the actual charge disproportionation is small. Paudel et al.[56] also studied the intrinsic defects in bulk $BiFeO_3$ and their effect on magnetization using first-principles approach. They found that the dominant defects in oxidizing (oxygen-rich) conditions are Bi and Fe vacancies and in reducing (oxygen-poor) conditions are O and Bi vacancies. The calculated carrier concentration shows that the $BiFeO_3$ grown in oxidizing conditions has p-type conductivity. The conductivity is reported to decrease with decreasing oxygen partial pressure, and the material shows insulating behaviour or n-type conductivity. According to these calculations, the Bi and Fe vacancies produce a magnetic moment of ~1$\mu_B$ and 5$\mu_B$ per vacancy, respectively, for p-type $BiFeO_3$ and none for insulating $BiFeO_3$. O vacancies do not introduce any moment for both p-type and insulating $BiFeO_3$. Since our samples are sintered in close atmosphere, so there is a possibility that the $BiFeO_3$ becomes n-type due to oxygen vacancies. Mn-doping is known to reduce the dielectric losses, increase the dc resistivity and improve the magnetization behaviour of $BiFeO_3$.[34,48-53] In case of our samples, the resistivity of undoped $BiFeO_3$ measured at room temperature is $1.6\times10^7\Omega$ cm while for the 0.3wt% $MnO_2$-doped $BiFeO_3$, the resistivity increases by three orders of magnitude to $1.09\times10^{10}\Omega$ cm which is comparable to that reported by Kumar et al. in Mn substituted $BiFeO_3$.[53] In $MnO_2$ doped $BiFeO_3$ sample, the higher resistivity is expected due to donor doping of Mn and reduced oxygen vacancy concentration. $Mn^{4+}$ ion plays a role of donor in $BiFeO_3$ because it possesses a higher valence than $Fe^{3+}$. The addition of $Mn^{4+}$ to $BiFeO_3$ requires charge compensation, which can be achieved by redox reaction involving capture of electron by $Mn^{4+}$ ion reducing it to $Mn^{3+}$. Further, the Mn ion is a multi- valence ion ($Mn^{2+,}$ $Mn^{3+}$ and $Mn^{4+)}$ which can be oxidated or reduced during the sintering processes. However, it has been reported[53] that $Mn^{4+}$ ion is not stable at high temperature and only $Mn^{3+}$ and $Mn^{2+}$ ions are stable in the ceramics at the sintering temperatures. Thus the following redox reaction can take place:[52]

$$Mn^{3+} + Fe^{2+} \rightarrow Fe^{3+} + Mn^{2+}$$

This reaction can effectively suppress the conversion of $Fe^{3+}$ to $Fe^{2+}$ and reduce the n-type doping effect due to oxygen vacancies leading to the observed increase in the resistivity of doped samples. According to the Shannon et al., the $Mn^{3+}$ (ionic radius r = 0 .645 Å) ion can occupy the $Fe^{3+}$ (r = 0 .645 Å) sites in $BiFeO_3$ materials, because both have the same valence state and ionic radius. So irrespective of the current level of understanding of the defect induced magnetism based on first principle calculations, $MnO_2$ doping does reduce the ionic vacancy concentrations and this may be responsible for the suppression of the 50 K cusp in the ZFC M (T) in the doped samples. Based on this discussion, we conclude that due to 0.3 wt% $MnO_2$ doping with reduced ionic vacancy concentrations, the intrinsic nature of the low temperature phase transitions in $BiFeO_3$ gets revealed unambiguously.

**B. Anomalous AC Susceptibility Response of $BiFeO_3$:**

Having established the existence of spin glass freezing around 25K in $BiFeO_3$ as an intrinsic feature, we now turn to some intriguing aspects of AC $\chi$ ($\omega$, T). Firstly, the peak height of $\chi'(\omega, T)$ increases with increasing frequencies which is unusual as the susceptibility always decreases with increasing frequency except near frequencies corresponding to a resonant absorption that may be linked with the resistance, capacitance and inductance of the entire circuit rather than just the sample. In principle, it is possible to push the resonance frequencies to higher side by reducing the capacitance and inductance of the circuit by reducing the sample quantity. However, this was not possible in $BiFeO_3$ due to very weak signal for the χ'(ω, T). A similar anomalous AC magnetic susceptibility data has been reported in single crystal samples of $BiFeO_3$.[17] It is also important to note that the imaginary part χ"(ω, T) shows negative cusps at $T_f$ with a peak temperature above the corresponding peak temperature for the real part χ'(ω, T). The negative value clearly suggests that the circuit is no longer purely inductive except at very low temperatures (< ~10K). The third intriguing aspect of the χ"(ω, T) is the presence of a tiny peak around 10K below which the imaginary part shows positive value. All these features require further study, as some of these anomalous features have also been tentatively attributed to the coexisting modulated magnetic structure of $BiFeO_3$,[15,16,57-61] not observed in the conventional spin glass systems. Further, the occurrence of a spin glass phase in a homogeneously ordered system like $BiFeO_3$ without any quenched impurity and randomness requires further investigation as the existing models of spin glass transitions are based on the concept of disorder, randomness and frustration.[32]

**C. Ground State of $BiFeO_3$:**

$BiFeO_3$ shows non-collinear AFM ordering with Heisenberg spins with an incommensurately modulated cycloidal spin structure superimposed on it. As said earlier, recent neutron scattering[15,16,57-61] and NMR studies[11-14] suggest that this spin cycloid is stable down to the lowest temperature (~5K). Considering these observations in conjunction with the present results, the most likely scenario for the ground state of $BiFeO_3$ is the coexistence of the spin glass phase and the long range ordered spin cycloid. The coexistence of LRO AFM and spin glass state has been a subject of extensive theoretical and experimental investigations for both Ising and Heisenberg systems.[62-65] In some of the Heisenberg systems, it has been predicted theoretically[62] and verified experimentally[63,64] that the coexistence is due to the freezing of the transverse component of the spins.[65] An alternative proposal in disordered systems like $PbFe_{0.5}Nb_{0.5}O_3$ (PFN) whereby the two phases result from two different magnetic sublattices, one (LRO) with percolative path ways and the other due to isolated Fe-O-Fe clusters, has also been proposed.[66] Pure $BiFeO_3$ has no quenched disorder per say, except for the possibility of $Fe^{2+}$ ions in the magnetic sublattice replacing some of the $Fe^{3+}$ sites due to redox reaction caused by oxygen vacancies and raising the possibility of local ferromagnetic interactions via double exchange. However, even though the oxygen vacancy concentrations, and hence the proportion of $Fe^{2+}$ and $Fe^{3+}$ in the magnetic sublattice, are expected to be significantly different in our undoped and doped $BiFeO_3$, the spin glass phase occurs below the same temperature $T_f$ with similar spin glass transition temperatures $T_{SG}$ and activation energies $E_{act}$. This indirectly suggests that the oxygen vacancies do not significantly influence the spin glass phase. In the absence of disorder in the magnetic sublattice or any frustrated interaction between the spins, the most likely possibility for the emergence of the spin glass phase is due to the freezing of the transverse component of the spins. Local magnetic probe like NMR[11-14] and the global probes like neutron scattering[15,16] have revealed the possibility of distortions in the long range ordered magnetic modulated structure even though the extent of distortion from harmonic modulation is much less for the average structure, probed by the bulk techniques like neutron scattering, than that reported by local probe like NMR. Whether this anhormonicity is linked with the gradual transverse freezing of the spins or not needs further investigation using neutron scattering studies on single crystals.

## V. CONCLUSIONS

The DC magnetization and AC susceptibility measurements on pure $BiFeO_3$ and 0.3 wt% $MnO_2$ doped $BiFeO_3$ show the existence of spin glass freezing around 25 K with a spin glass transition temperature $T_{SG}$ ~20 K. The anomaly around 50 K could be due to extrinsic factors like oxygen vacancies as it is not present in the $MnO_2$ doped samples where the vacancy concentration is drastically reduced. The two other anomalies seen in the M (T) of the doped sample seem to suggest that 250 K and 100-150 K transitions are also intrinsic to $BiFeO_3$.

Combining the recent neutron and NMR studies on the presence of long range ordered magnetic phase at low temperatures (upto ~ 5K) and the present results, the most likely scenario for the ground state of $BiFeO_3$ is the coexistence of spin glass phase with the long range ordered spin cycloid with somewhat anhormonic modulation.

## SUPPLEMENTRY MATERIALS

See supplementary file: EDX analysis, LeBail profile fitting, refined unit cell parameters at room temperature and Power-law analysis for pure and 0.3 wt% $MnO_2$ doped $BiFeO_3$.

## ACKNOWLEDGEMENTS

The authors thank Dr. Alok Banerjee, UGC-DAE Consortium for Scientific Research Indore, India for magnetization measurements. Dhananjai Pandey acknowledges financial support from Science and Engineering Research Board (SERB) of India through J. C. Bose Fellowship grant. The authors also acknowledge to Girish Sahu, CIF, IIT (BHU) Varanasi for EDX analysis.

TABLE I. Comparison of fitting parameters including goodness of fit (GoF) for the two models for pure and 0.3 wt% $MnO_2$ doped $BiFeO_3$.

| Vogel-Fulcher Law | | | Scaling Law | | |
|---|---|---|---|---|---|
| Parameters | Pure $BiFeO_3$ | Doped $BiFeO_3$ | Parameters | Pure $BiFeO_3$ | Doped $BiFeO_3$ |
| $T_{VF}$ | 21.61 K | 20.07 K | $z\nu$ | 1.862 | 1.582 |
| $E_{act}$ | 0.822 meV | 0.849 meV | $T_{SG}$ | 21.61 K | 20.07 K |
| $\tau_0$ | 6.515 x $10^{-5}$ s | 9.959 x $10^{-5}$ s | $\tau_0$ | 2.975 x $10^{-5}$ s | 7.922 x $10^{-5}$ s |
| GoF | 0.99877 | 0.99912 | GoF | 0.99455 | 0.99576 |

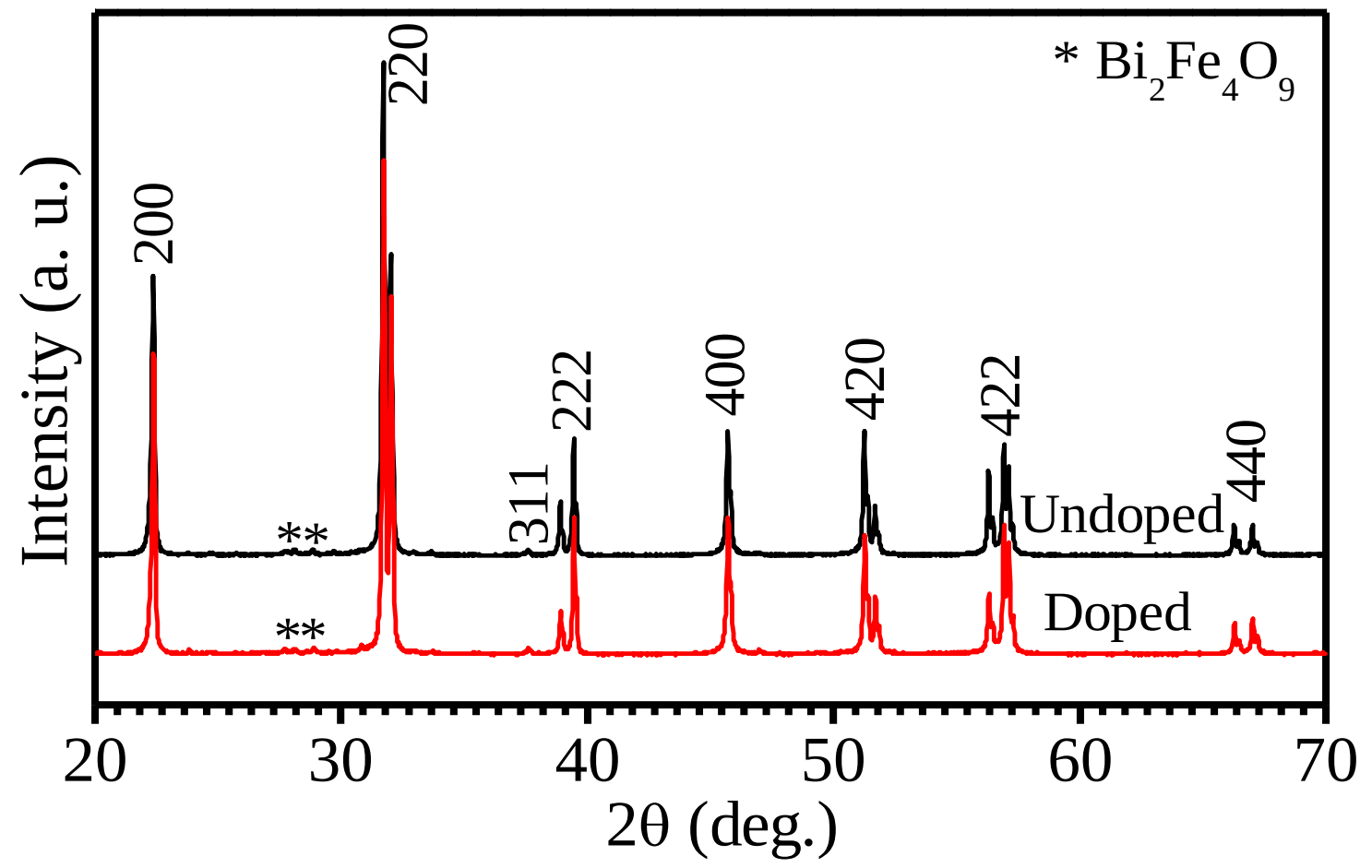

**FIG. 1.** X-ray diffraction patterns of pure $BiFeO_3$ and 0.3 wt% $MnO_2$ doped $BiFeO_3$ collected at room temperature. All indices are with respect to a doubled pseudocubic cell. The asterisk (*) marks the impurity peak of $Bi_2Fe_4O_9$.

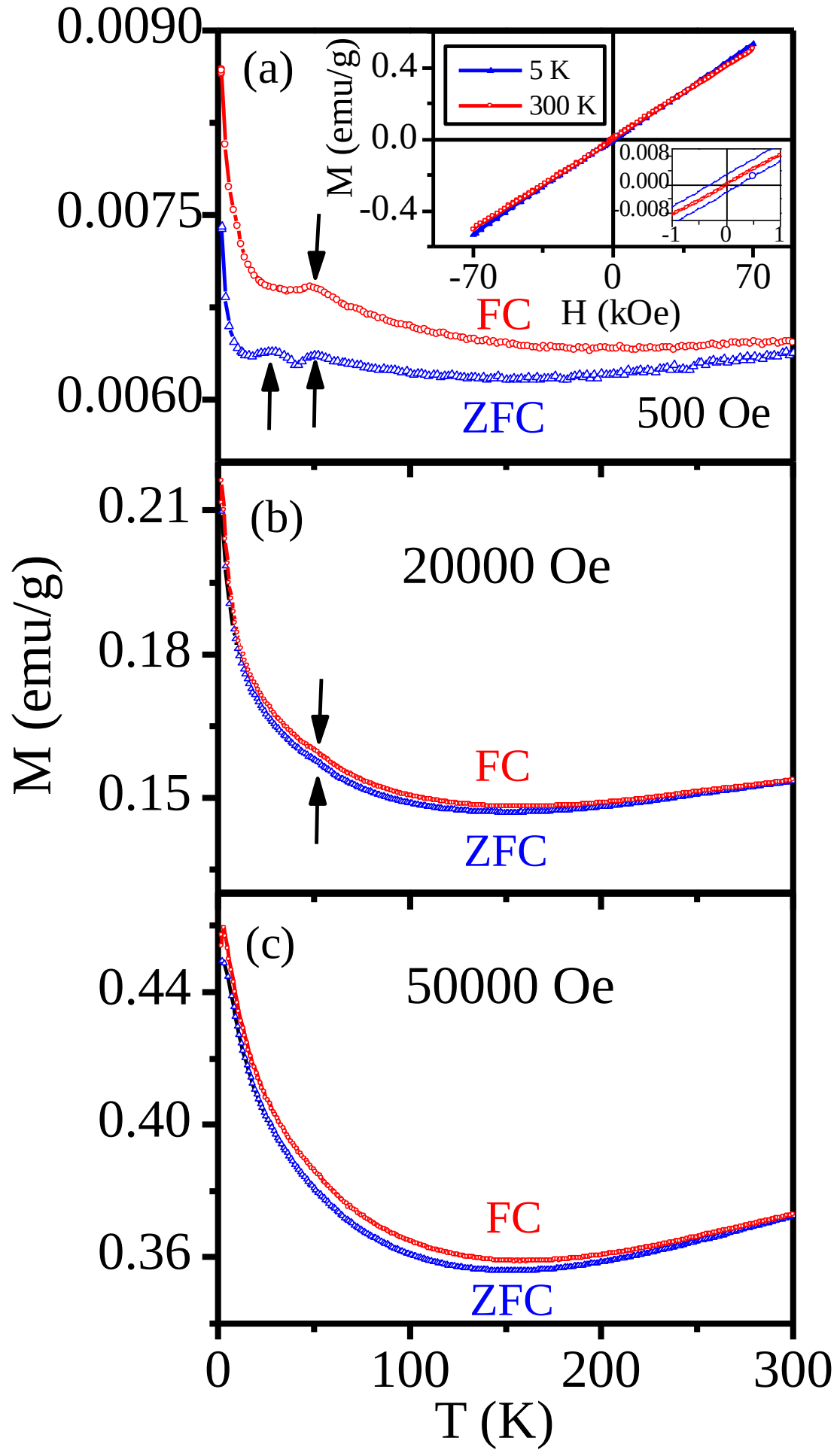

**FIG. 2.** Temperature dependence of dc magnetization (M) at an applied dc field of **(a)** 500 Oe **(b)** 20000 Oe and **(c)** 50000 Oe for pure $BiFeO_3$. Insets to figure depict the M-H plots at 5 K and 300 K.

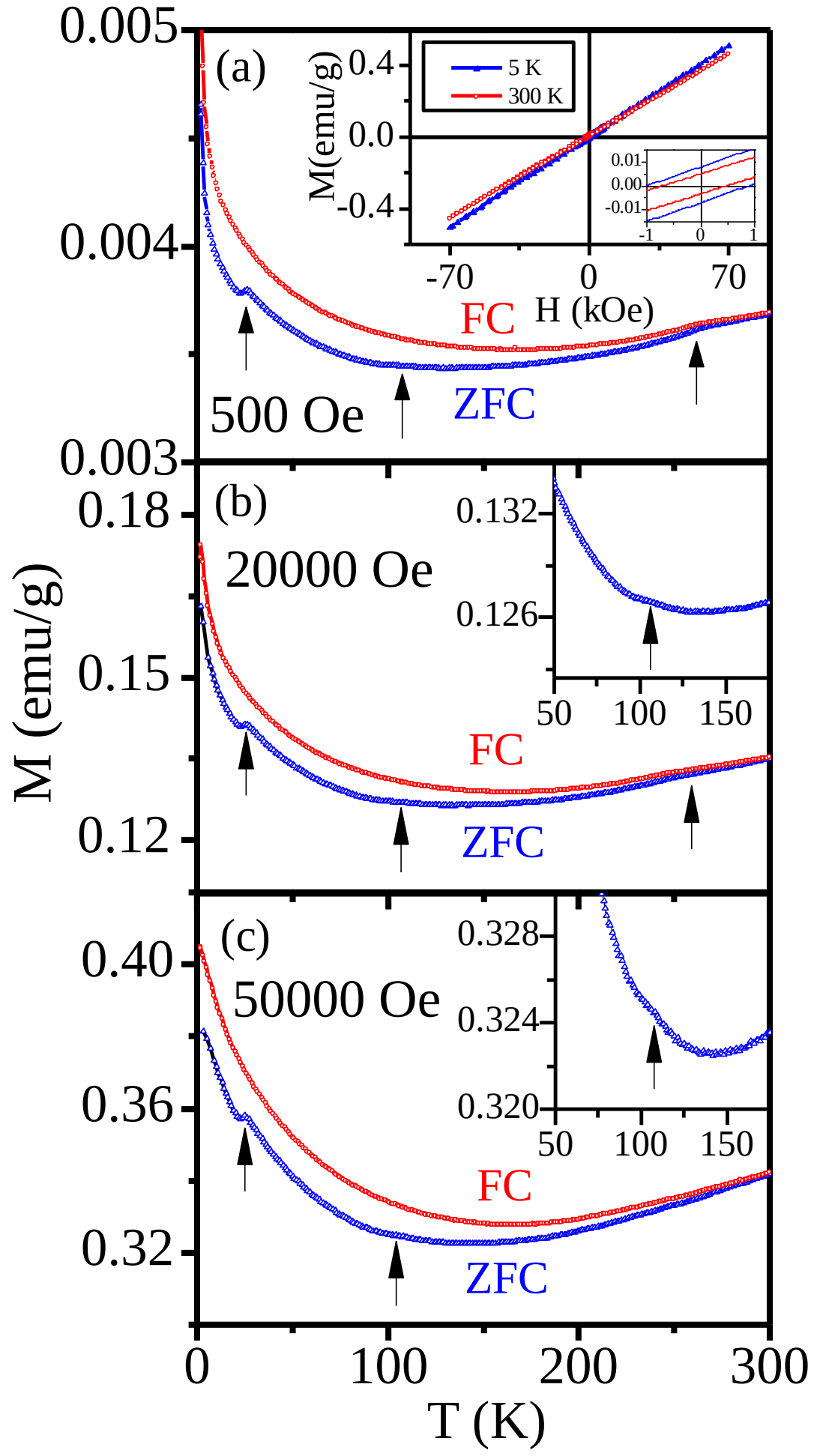

**FIG. 3.** Temperature dependence of dc magnetization (M) at an applied dc field of **(a)** 500 Oe **(b)** 20000 Oe and **(c)** 50000 Oe for 0.3 wt% $MnO_2$ doped $BiFeO_3$. Insets to figure depict the M-H plots at 5 K and 300 K.

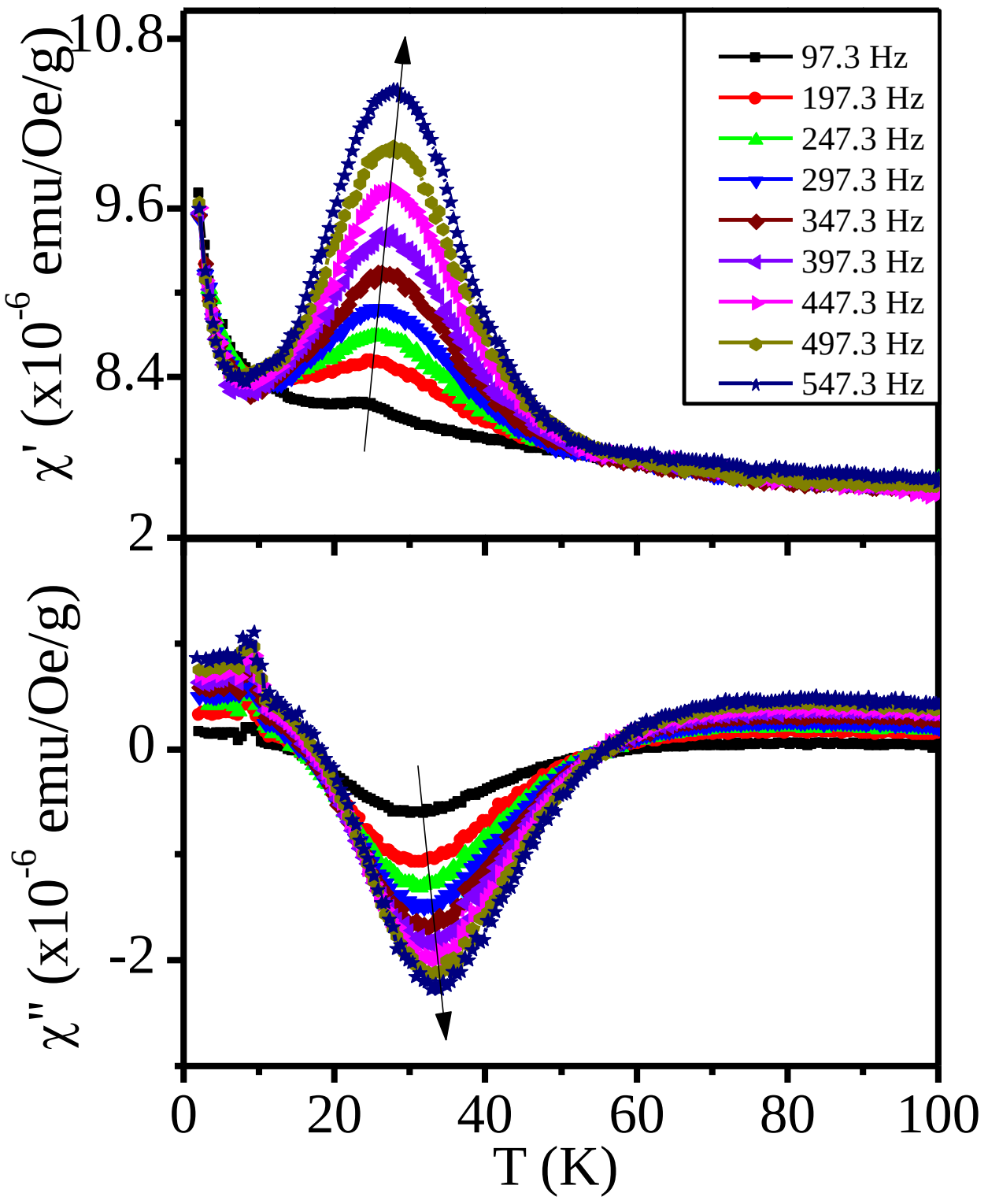

**FIG. 4.** Temperature dependence of the real and imaginary parts of the ac susceptibility at various frequencies at an applied ac field of 5 Oe for pure $BiFeO_3$.

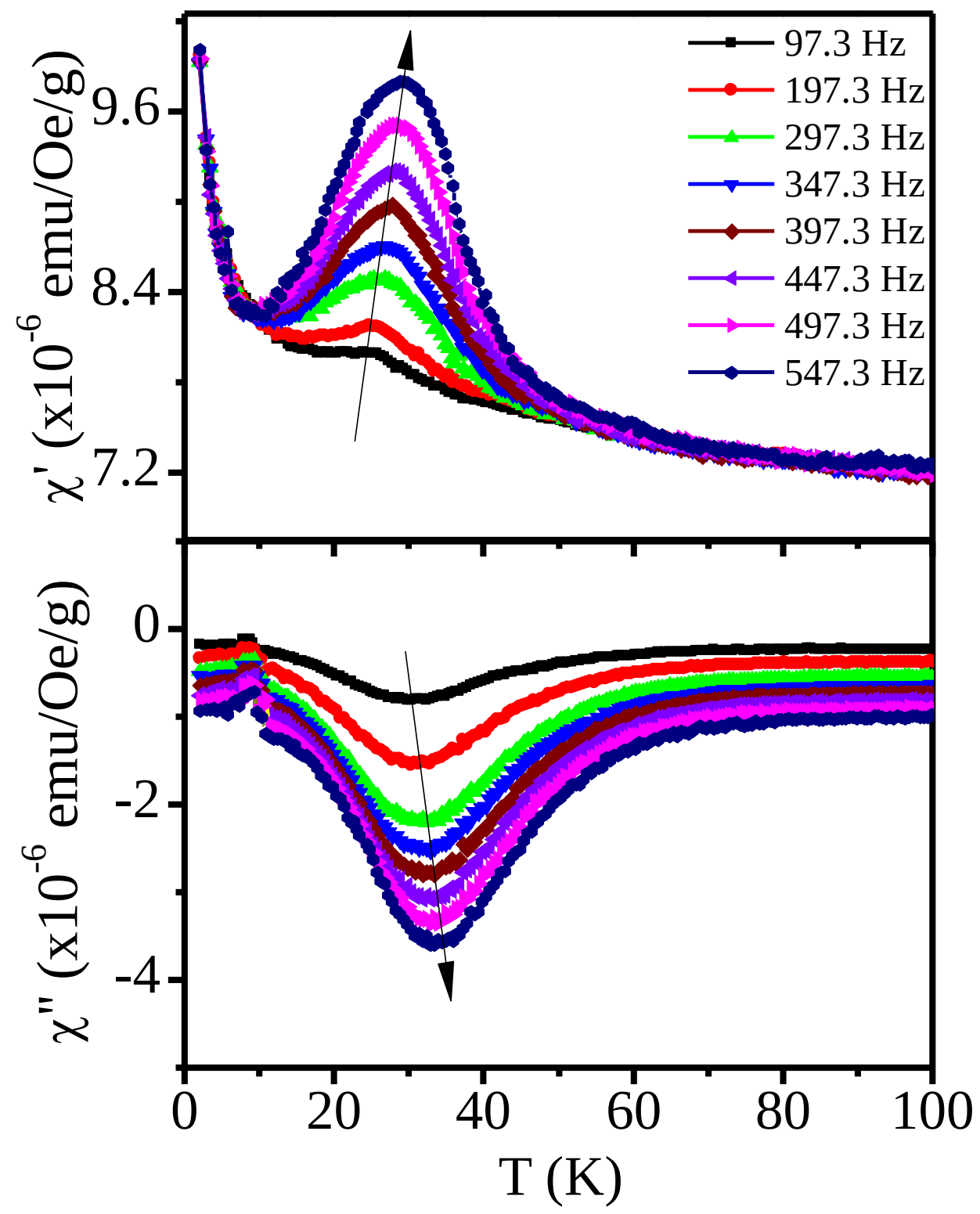

**FIG. 5.** Temperature dependence of the real and imaginary parts of the ac susceptibility at various frequencies at an applied ac field of 5 Oe for 0.3 wt% $MnO_2$ doped $BiFeO_3$.

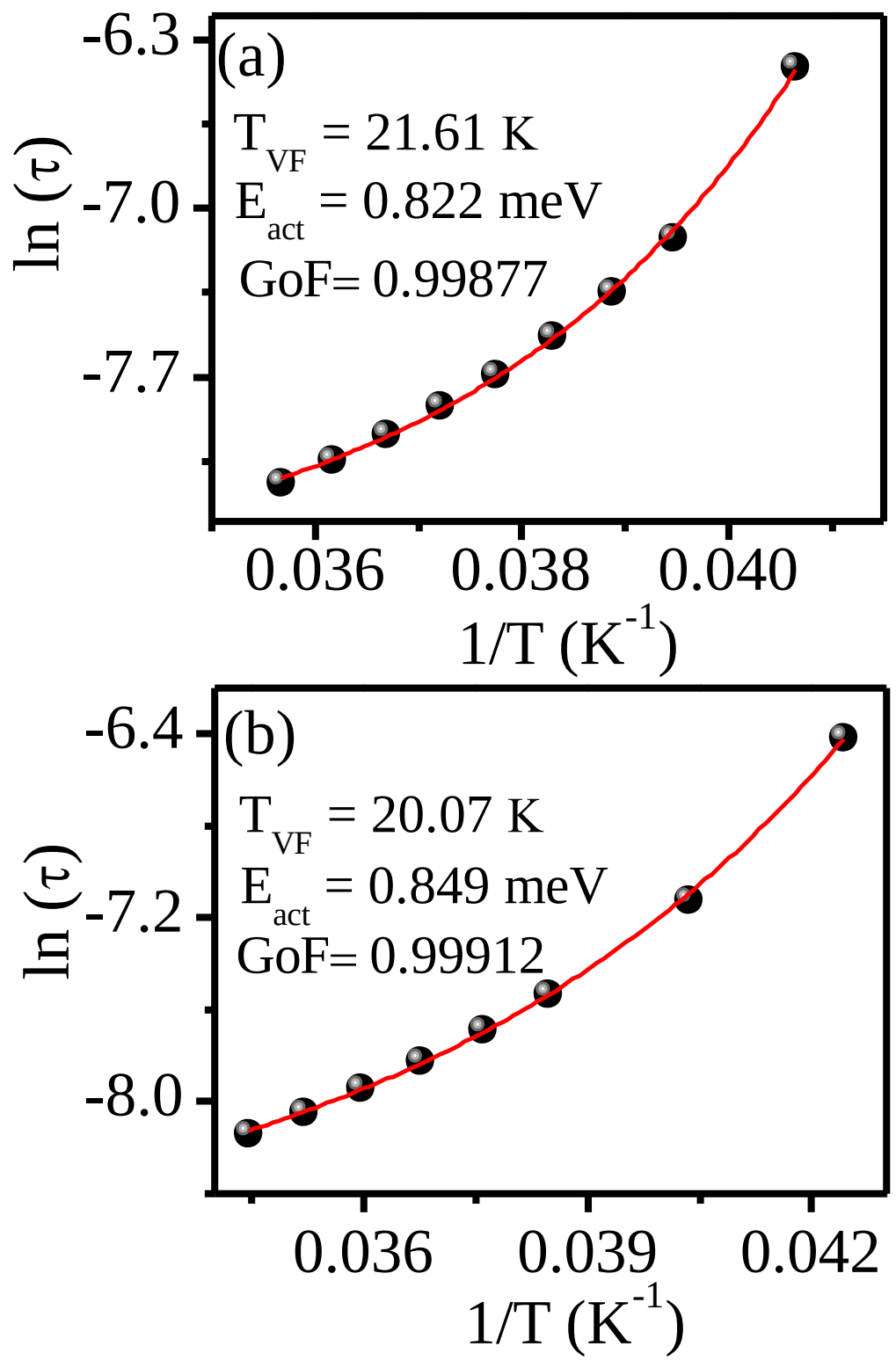

**FIG. 6.** lnτ vs 1/T plot. The solid line is the fit for the Vogel-Fulcher law for **(a)** pure $BiFeO_3$ and **(b)** 0.3 wt% $MnO_2$ doped $BiFeO_3$.

# Supplementary File

## 1. Composition analysis:

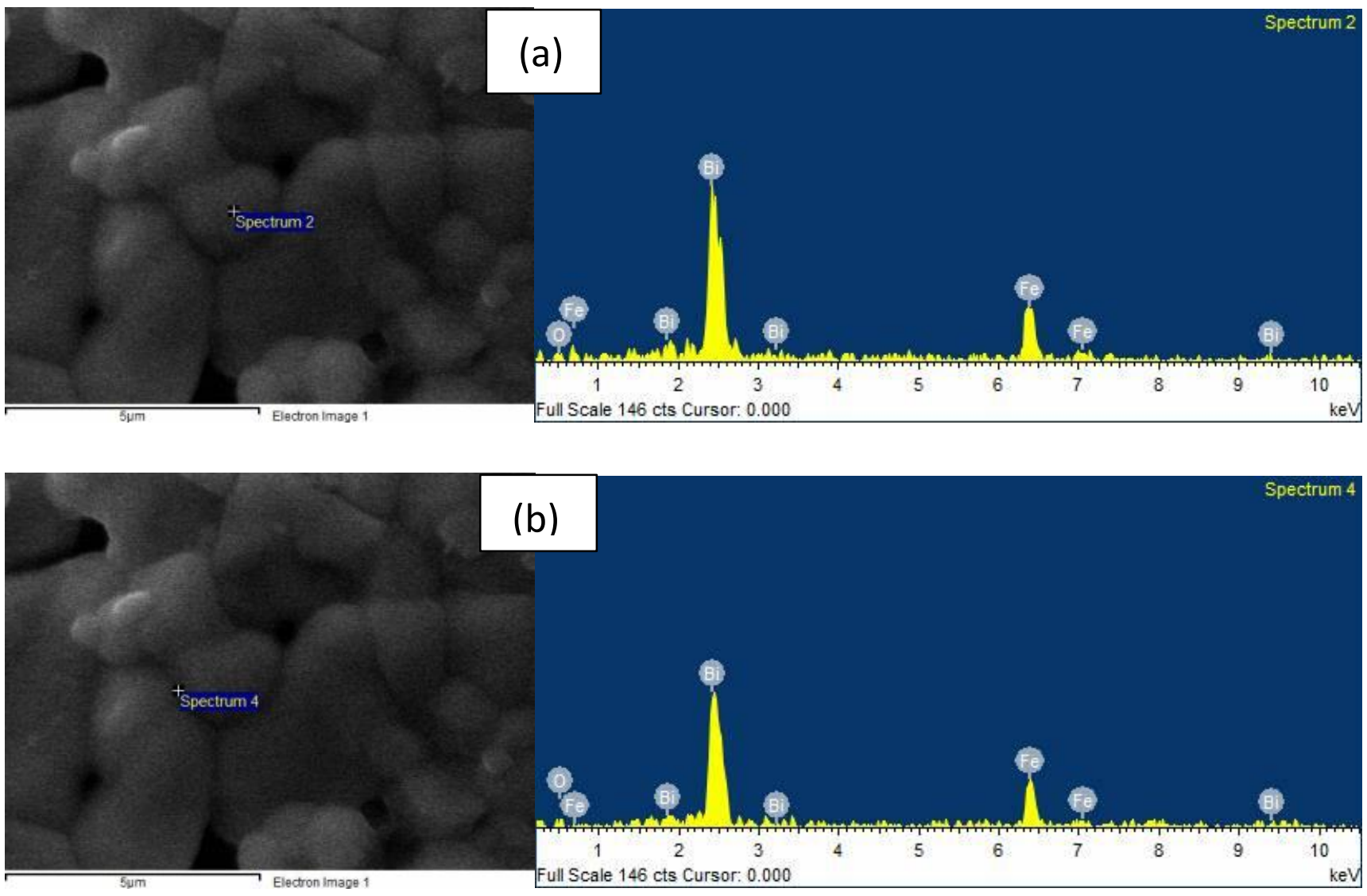

**Fig. S1.** Microstructure and EDX spectra of $BiFeO_3$ **(a)** in the grain and **(b)** at the grain boundary

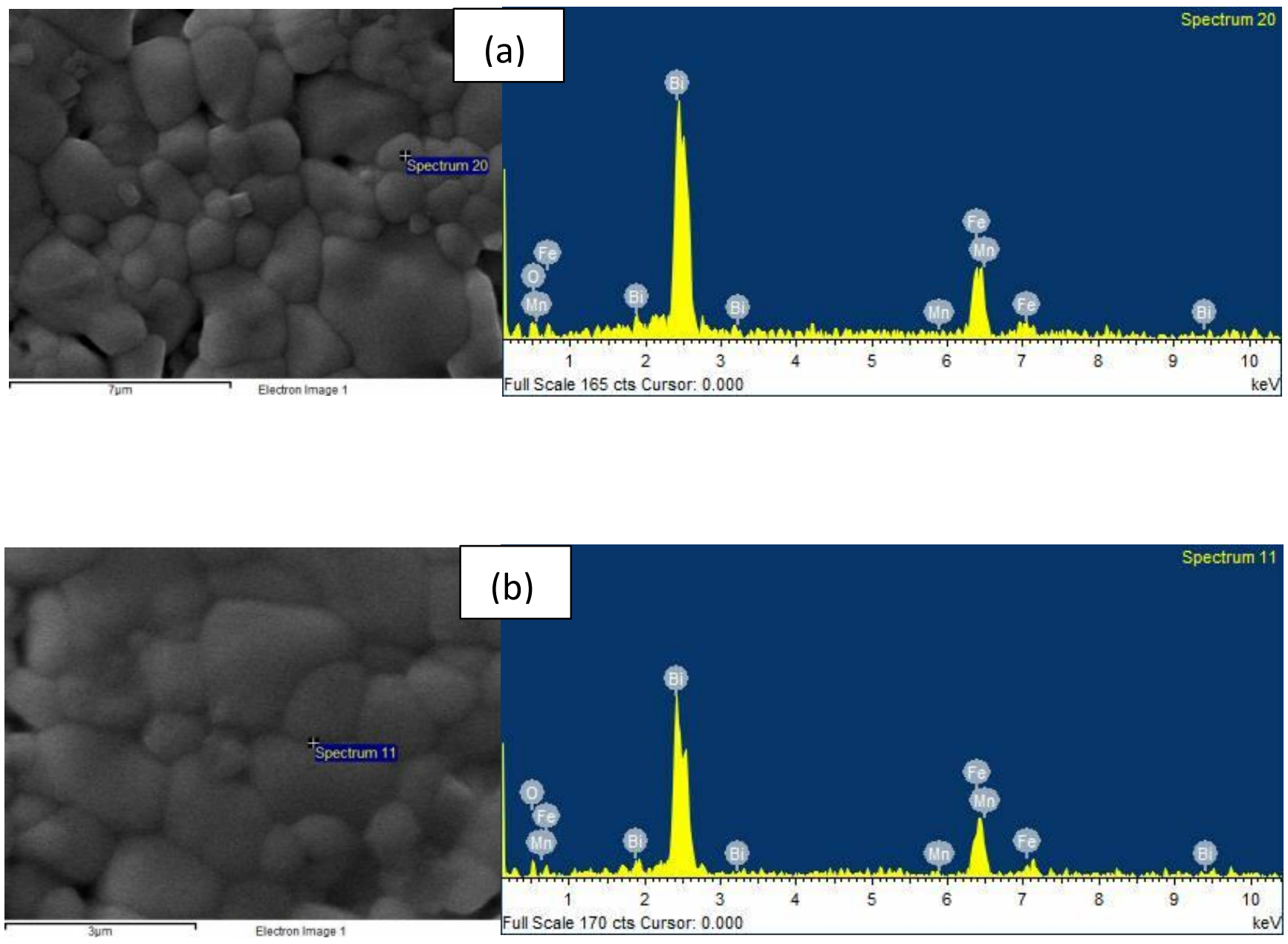

**Fig. S2.** Microstructure and EDX spectra of 0.3% $MnO_2$ doped $BiFeO_3$ **(a)** in the grain and **(b)** at the grain boundary.

**Table S1:** Results of EDX analysis of pure $BiFeO_3$ and 0.3wt % $MnO_2$ doped $BiFeO_3$ in weight percent for the microstructure and spectra shown in Fig. S1 and S2.

| Pure $BiFeO_3$ | | | 0.3wt% $MnO_2$ doped $BiFeO_3$ | | |
|---|---|---|---|---|---|
| Element | Weight % | | Element | Weight % | |
| | Grain | Grain Boundary | | Grain | Grain Boundary |
| O | 11.90 | 11.28 | O | 12.54 | 14.54 |
| Fe | 18.74 | 18.75 | Mn | 0.33 | 0.34 |
| Bi | 69.36 | 69.97 | Fe | 17.58 | 18.50 |
| | | | Bi | 69.54 | 66.62 |
| Total | 100.00 | 100.00 | Total | 100.00 | 100.00 |

**Table S2:** Average composition of the pure $BiFeO_3$ and 0.3wt % $MnO_2$ doped $BiFeO_3$ samples in weight percent.

| Pure $BiFeO_3$ | | | 0.3wt% $MnO_2$ doped $BiFeO_3$ | | |
|---|---|---|---|---|---|
| Element | Weight% | | Element | Weight % | |
| | Expected | Average | | Expected | Average |
| O | 15.3 | 13.0 ± 1.7 | O | 15.3 | 13.6 ± 1.4 |
| Fe | 17.9 | 18.6 ± 0.3 | Mn | 0.30 | 0.3 ± 0.1 |
| Bi | 66.8 | 68.4 ± 1.5 | Fe | 17.9 | 17.8 ± 0.7 |
| | | | Bi | 66.8 | 68.3 ± 1.3 |
| Total | 100.00 | 100.00 | Total | 100.00 | 100.00 |

## 2. LeBail refinement:

The LeBail refinement using R3c space group of $BiFeO_3$ [7] was carried out for both the samples using FULLPROF package [35]. The observed (filled-circles) and calculated (continuous line) profiles show excellent fit for both the samples, as can be seen from the difference (bottom line) profile given in Fig. S3. This confirms that both the samples belong to the R3c space group. The refined unit cell parameters are listed in Table S3.

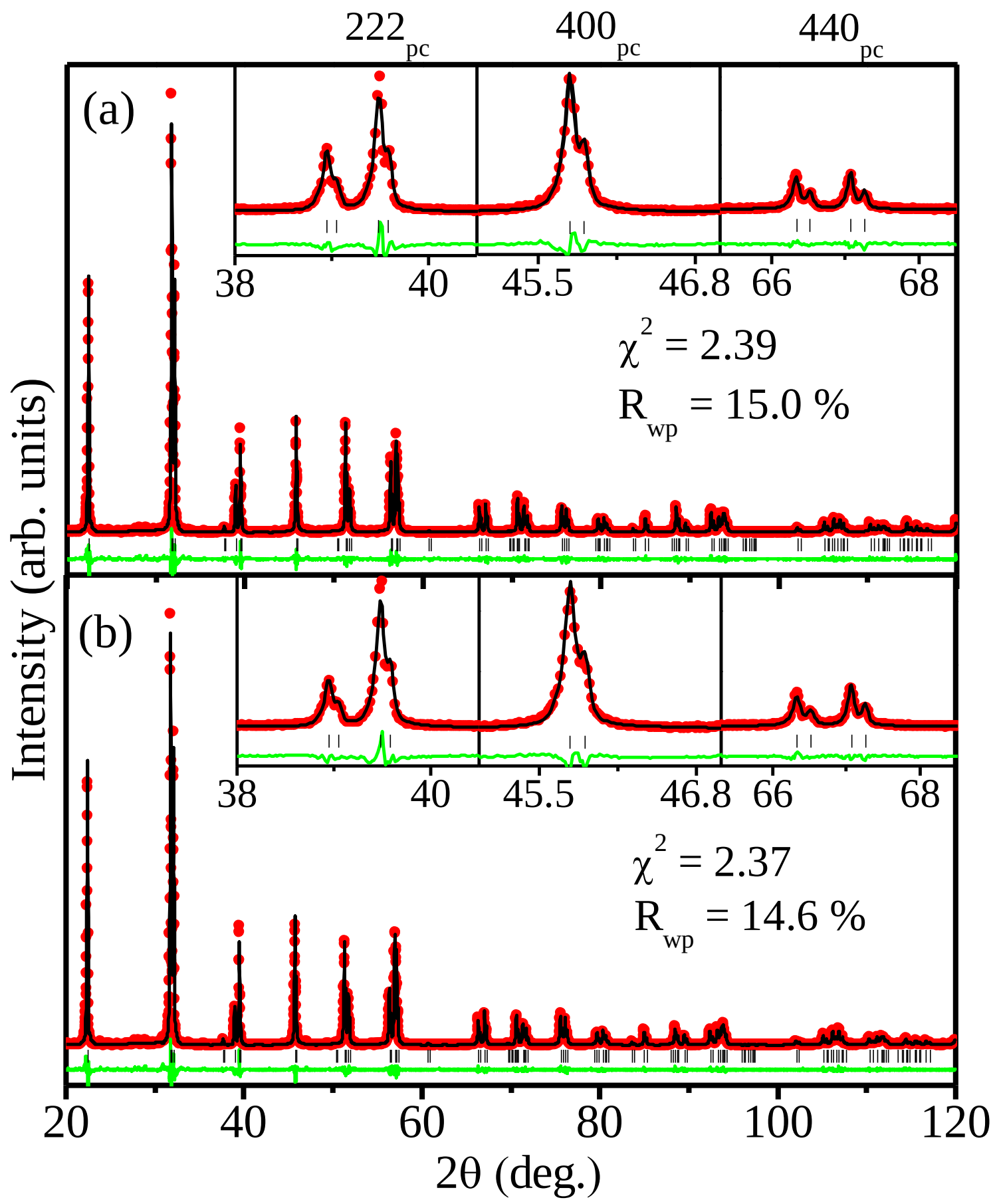

**Fig. S3.** Observed (filled circles), calculated (continuous line), and difference (bottom line) profiles obtained from LeBail refinement at room temperature using R3c space group for **(a)** pure $BiFeO_3$ and **(b)** 0.3 wt% $MnO_2$ doped $BiFeO_3$.

**Table S3:** LeBail refined unit cell parameters for pure $BiFeO_3$ and 0.3 wt% $MnO_2$ doped $BiFeO_3$ at room temperature.

| Parameters | Pure $BiFeO_3$ | 0.3 wt% $MnO_2$ doped $BiFeO_3$ |
|---|---|---|
| Space group | R3c<br>Hexagonal unit cell parameters a = b ≠ c, α = β = $90^0$, γ = $120^0$ | |
| a (Å) | 5.5772 (6) | 5.5775 (7) |
| c (Å) | 13.8648 (1) | 13.8653 (2) |
| V ($Å^3$) | 373.501 (8) | 373.552 (9) |
| $R_{wp}$ (%) | 15.0 | 14.6 |
| $\chi^2$ | 2.39 | 2.37 |

### 3. Spin glass dynamics using power law:

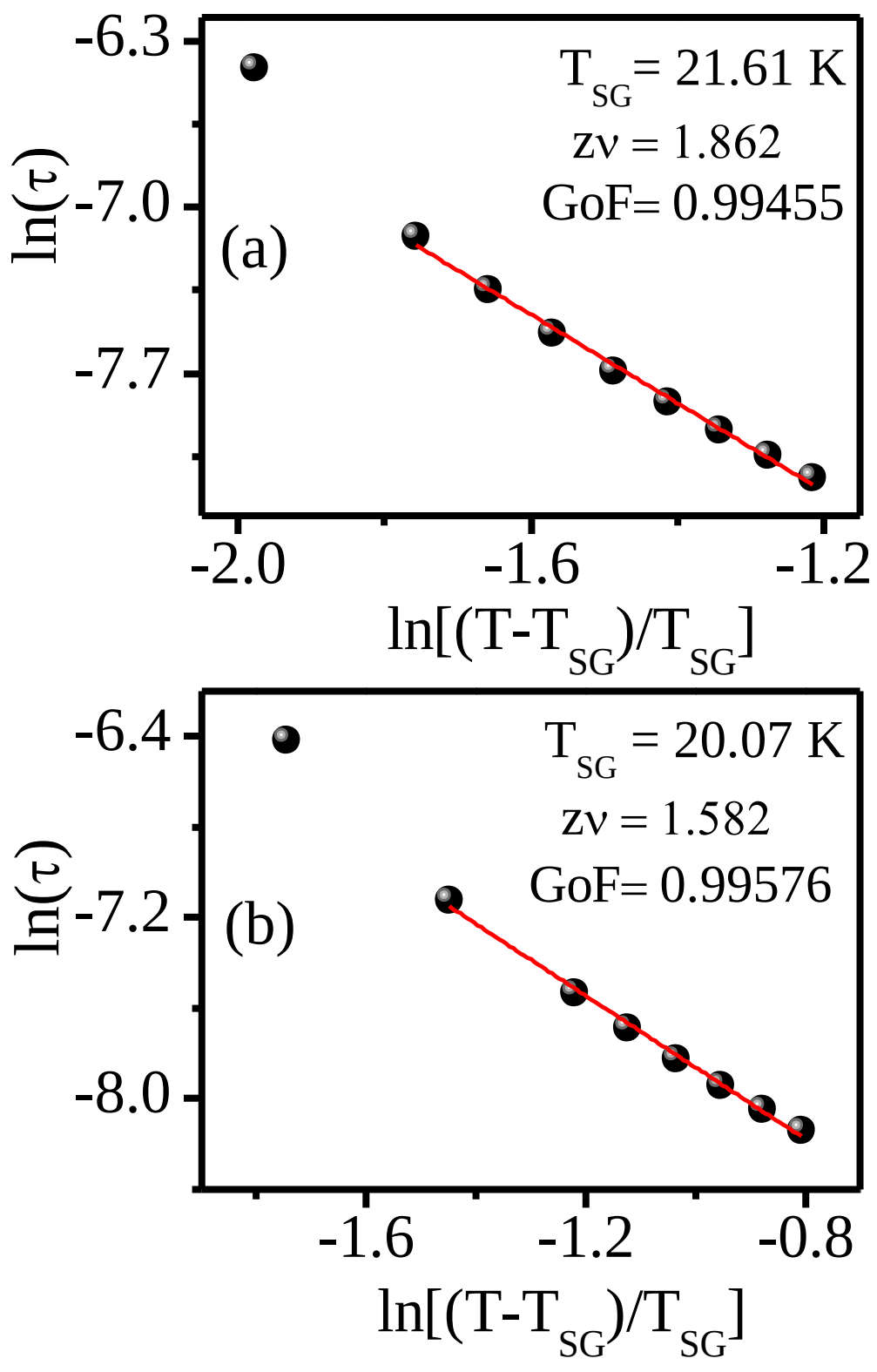

**Fig. S4.** lnτ vs ln(T-$T_{SG}$/$T_{SG}$) plot. The solid line is the fit for the power law for **(a)** pure $BiFeO_3$ **(b)** 0.3 wt% $MnO_2$ doped $BiFeO_3$.